\documentclass[12pt]{iopart}

\usepackage{iopams, float}
\usepackage{cite}
\usepackage{graphicx,caption}
\usepackage{lmodern}
\usepackage{mathrsfs}
\usepackage{amsfonts}
\usepackage[utf8]{inputenc}
\usepackage{url}
\usepackage[colorlinks]{hyperref}
\usepackage[dvipsnames]{xcolor}
\usepackage{multirow}
\usepackage[normalem]{ulem}
\usepackage{lipsum}

\usepackage{marvosym}
\usepackage{enumerate}
\usepackage{color,soul}

\usepackage{bm}
\usepackage{color}
\usepackage{multirow}
\usepackage{acronym}

\newacro{adm}[ADM]{Arnowitt-Deser-Misner}
\newacro{bbh}[BBH]{binary black hole}
\newacro{bh}[BH]{black hole}
\newacroplural{bh}[BHs]{black holes}
\newacro{bhns}[BHNS]{black hole-neutron star}
\newacro{bhpt}[BHPT]{black hole perturbation theory}
\newacro{bns}[BNS]{binary neutron star}
\newacro{bf}[BF]{Bayes' factor}
\newacro{cbc}[CBC]{compact binary coalescence}
\newacro{ce}[CE]{Cosmic Explorer}
\newacro{da}[DA]{data analysis}
\newacro{et}[ET]{Einstein Telescope}
\newacro{eob}[EOB]{Effective-One-Body}
\newacro{eom}[EOM]{equations of motion}
\newacro{fd}[FD]{frequency domain}
\newacro{fft}[FFT]{Fast Fourier transform}
\newacro{gw}[GW]{gravitational-wave}
\newacroplural{gw}[GWs]{Gravitational-waves}
\newacro{gr}[GR]{general relativity}
\newacro{grb}[GRB]{gamma-ray burst}
\newacro{grhd}[GRHD]{general-relativistic hydrodynamics}
\newacro{gwosc}[GWOSC]{Gravitational Wave Open Science Center}
\newacro{gwtc1}[GWTC-1]{the first gravitational-wave transients catalog}
\newacro{gsf}[GSF]{Gravitational Self Force}
\newacro{hm}[HM]{Higher mode}
\newacroplural{hm}[HMs]{Higher modes}
\newacro{ifo}[IFO]{interferometer}
\newacro{imr}[IMR]{inspiral-merger-ringdown}
\newacro{im}[IMR]{inspiral-to-merger}
\newacro{kagra}[KAGRA]{Kamioka Gravitational Wave Detector}
\newacro{ligo}[LIGO]{Laser Interferometer Gravitational-Wave Observatory}
\newacro{lisa}[LISA]{Laser Interferometer Space Antenna}
\newacro{lr}[LR]{Light Ring}
\newacro{lso}[LSO]{Last Stable Orbit}
\newacro{lvc}[LVC]{LIGO-Virgo Collaboration}
\newacro{lvk}[LVK]{LIGO-Virgo-Kagra Collaboration}
\newacro{lo}[LO]{leading order}
\newacro{ns}[NS]{neutron star}
\newacroplural{ns}[NSs]{neutron stars}
\newacro{nr}[NR]{numerical relativity}
\newacro{nqc}[NQCs]{Next-to-quasicircular corrections}
\newacro{nlo}[NLO]{next-to-leading order}
\newacro{nnlo}[NNLO]{next-to-next-to-leading order}
\newacro{n3lo}[N3LO]{next-to-next-to-next-to-leading order}
\newacro{n4lo}[N3LO]{next-to-next-to-next-to-next-to-leading order}
\newacro{ode}[ODE]{Ordinary Differential Equation}
\newacroplural{ode}[ODEs]{Ordinary Differential Equations}
\newacro{pn}[PN]{post-Newtonian}
\newacro{pm}[PM]{post-Minkowskian}
\newacro{pe}[PE]{parameter estimation}
\newacro{psd}[PSD]{power spectral density}
\newacro{pa}[PA]{post-adiabatic}
\newacro{qnm}[QNM]{quasi-normal mode}
\newacro{qc}[QC]{quasi-circular}
\newacro{snr}[SNR]{signal-to-noise ratio}
\newacro{spa}[SPA]{stationary-phase approximation}
\newacro{sxs}[SXS]{Simulating eXtreme Spacetimes}
\newacro{td}[TD]{time domain}
\newacro{ng}[NG]{Nect Generation}
\newacro{rift}[\texttt{RIFT}]{Rapid Iterative FiTting}

\def\gw{GW150914}
\def\TEOBResumS{\texttt{TEOBResumS-Dal\'i}}

\definecolor{cyan}{rgb}{0,0.9,0.9}
\definecolor{orange}{rgb}{0.9,0.5,0}
\definecolor{magenta}{rgb}{1,0,1}
\definecolor{purple}{rgb}{0.8,0.4,0.8}
\definecolor{gray}{rgb}{0.8242,0.8242,0.8242}
\definecolor{dodgerblue}{rgb}{0.12, 0.56, 1.0}

\begin{document}

\title{Revisiting GW150914 with a non-planar, eccentric waveform model}
\author{Rossella Gamba${}^{1,2,3, *}$, Jacob Lange${}^{4}$, Danilo Chiaramello${}^{4,5}$, Jacopo Tissino${}^{6,7}$ and Snehal Tibrewal${}^{8}$}
\address{${}^{1}$ Institute for Gravitation and the Cosmos, The Pennsylvania State University, University Park, PA 16802, USA}
\address{${}^{2}$ Department of Physics, The Pennsylvania State University, University Park, PA 16802, USA}
\address{${}^{3}$ Department of Physics, University of California, Berkeley, CA 94720, USA}
\address{${}^{4}$ INFN, Sezione di Torino, Via Pietro Giuria 1, I-10125 Torino, Italy}
\address{${}^{5}$ Universit\`a di Torino, Dipartimento di Fisica, Via Pietro Giuria 1, I-10125 Torino, Italy}
\address{${}^{6}$ Gran Sasso Science Institute (GSSI), I-67100 L'Aquila, Italy}
\address{${}^{7}$ INFN, Laboratori Nazionali del Gran Sasso, I-67100 Assergi, Italy}
\address{${}^{8}$ Center of Gravitational Physics, University of Texas at Austin, Austin, TX 78712, USA}
\address{$^*$Author to whom any correspondence should be addressed.}

\ead{rgamba@berkeley.edu}
\vspace{10pt}

\begin{indented}
\item[]\date{\today}
\end{indented}

\begin{abstract}
  The first direct detection of gravitational waves by the LIGO collaboration, GW150914, marked the start of a new exciting
  era in astronomy, enabling the study of the Universe through a new messenger. Since then, the field has grown rapidly, with
  the development of increasingly more sophisticated techniques to detect, analyze and interpret the signals.
  In this paper we revisit GW150914, presenting updated estimates of its source parameters using a 
  waveform model developed within the EOB formalism, able to describe gravitational-wave emission from generic non-circular,
  non-planar binaries. 
  We provide a comprehensive analysis of the signal and its properties, considering and contrasting various scenarios 
  for the source: from the simplest, aligned-spin quasi-circular binary black hole merger, to
  more complex scenarios, including precession, eccentricity or both.
  Unsurprisingly, we find that the signal is consistent with a quasi-circular ($e < 0.08$ at $15$ Hz), slowly spinning 
  $(\chi_{\rm eff} = -0.03^{+0.12}_{-0.13})$ binary black hole merger, a-posteriori validating a considerable body of works.
  This is the first analysis performed with an inspiral-merger-ringdown model containing both eccentricity and precession.
\end{abstract}

\submitto{\CQG}

\newpage

\section{Introduction}

On September 14 2015 the \ac{lvc}~\cite{KAGRA:2013rdx, LIGOScientific:2014pky, VIRGO:2014yos}
directly detected the first \ac{gw} signal from two compact objects, \gw~\cite{LIGOScientific:2016aoc}.
To date, this event represents one of the loudest, best
characterized \ac{gw} signals.
Since its discovery, \gw~has been the subject of numerous
studies across multiple domains of gravitational wave astronomy. 
Initial efforts were directed at characterizing the signal and its astrophysical properties~\cite{LIGOScientific:2016vlm,LIGOScientific:2016wkq}, 
using it to test the predictions of general relativity~\cite{LIGOScientific:2016lio} and
studying the properties of the \ac{bbh} population~\cite{LIGOScientific:2016kwr, LIGOScientific:2016vpg}.
These early analyses confirmed the consistency of \gw~with \ac{gr}, identifying the signal as the coalescence 
of two approximately equal-mass, slowly spinning \acp{bh} merging along quasi-circular orbits into a 
final Kerr \ac{bh} remnant with a mass of approximately $62 M_\odot$~\cite{LIGOScientific:2016wkq}.
A multitude of subsequent studies have utilized \gw~as a testbed for developing new techniques to analyze \ac{gw} signals. 
These include -- but are not limited to -- advancements in parameter estimation~\cite{Romero-Shaw:2020owr, Green:2020dnx, Breschi:2021wzr, Dax:2022pxd, Srinivasan:2024uax}, 
ringdown analyses~\cite{Carullo:2019flw, Carullo:2021yxh, Cotesta:2022pci, Isi:2022mhy, Correia:2023bfn, Gennari:2023gmx, Maenaut:2024oci, Wang:2024yhb, Pacilio:2024qcq}, 
new tests of GR~\cite{Carullo:2018gah, Isi:2020tac,Laghi:2020rgl, Carullo:2021oxn, Carullo:2021dui,Silva:2022srr}, 
and improvements to waveform models~\cite{Pratten:2020ceb, Riemenschneider:2021ppj, Gamba:2021ydi, Estelles:2020twz, Bonino:2022hkj, 
Ramos-Buades:2023ehm, Gamboa:2024hli}.
Despite the application of increasingly refined techniques, the fundamental interpretation 
of the signal has remained largely unchanged.

In this paper, we revisit \gw~presenting a comprehensive analysis of
its properties. We employ \TEOBResumS, a physically complete model based on the \ac{eob} framework~\cite{Gamba:2024cvy, Albanesi:2025txj},
to study the signal considering a variety of scenarios for the source: from the simple
coalescence of a planar, quasi-circular \ac{bbh} system, to more complex scenarios including
precession, eccentricity or, for the first time, the combination of the two.

The paper is structured as follows. In Sec.~\ref{sec:model} we summarize the main effects that
spin-induced precession and eccentricity have on a \ac{gw} signal from \ac{bbh} systems,
describing the techniques used to model them and their implementation within \TEOBResumS.
We then present a series of comparisons against selected \ac{nr} simulations.
In Sec.~\ref{sec:pe} we introduce the Bayesian \ac{pe} framework used to analyze \gw, and
describe the different hypotheses considered in our analysis.
Section~\ref{sec:results} collects the main results of our analysis, discussing the
posterior distributions of the parameters of interest and the implications for the
astrophysical interpretation of the signal.
Finally, we summarize our findings in Sec.~\ref{sec:conclusions}.

\paragraph{Conventions} We denote the masses of the two \acp{bh} as $m_1$ and $m_2$, their mass ratio as $q = m_1/m_2 \geq 1$ and their
symmetric mass ratio as $\nu = q/(1+q)^2$. The total mass is $M=m_1 + m_2$, while the chirp mass is $\mathcal{M} = (m_1 m_2)^{3/5}/(m_1 + m_2)^{1/5} = \nu^{3/5} M$.
We denote the dimensionless spin vectors of the two bodies as $\bm{\chi}_i = \bm{S}_i/m_i^2$, where $\bm{S}_i$ is the spin angular momentum of the $i$-th \ac{bh}.
The total angular momentum is then $\bm{J} = \bm{S}_1 + \bm{S}_2 + \bm{L}$, with $\bm{L}$ being the orbital angular momentum of the system.
We define the effective spin as $\chi_{\rm eff} = (m_1 \bm{\chi}_1 + m_2 \bm{\chi}_2) \cdot \hat{\bm{L}}/M$,
where $\hat{\bm{L}}$ is the unit vector along the direction of the orbital angular momentum, and the perpendicular spin parameter~\cite{Schmidt:2014iyl}
as:
\begin{equation}
  \chi_{\rm p} = \max \left\{|\bm{\chi}_1^{\perp}|, \frac{4 + 3q}{4q^2 + 3q} |\bm{\chi}_2^{\perp}|\right\} \, ,
\end{equation}
where $|\bm{\chi}_i^{\perp}|$ are the projections of the spin vectors onto the plane perpendicular to $\bm{L}$.
The waveform polarizations $h_+$, $h_\times$ are decomposed in terms of the $s = -2$ spin-weighted spherical harmonics ${}_{-2}Y_{lm}$ as
\begin{equation}
  h_+ - i h_\times = \sum_{lm} h_{lm} {}_{-2}Y_{lm}(\iota, \phi_{\rm ref}) ,
\end{equation}
where $\iota$ and $\phi_{\rm ref}$ are the angles that define the orientation of the line of sight 
with respect to the inertial reference system of the binary at the frequency $f_{\rm ref}$.
Unless otherwise specified, we use geometrized, mass-rescaled units in which $G = c = 1$ and
times are measured in multiples of the total mass $M$.

\section{Modeling GWs from BBHs on generic orbits}
\label{sec:model}

The accurate modeling of \acp{gw} emitted by coalescing \ac{bbh}
systems represents a significant challenge due to the complex interplay of relativistic effects.
Among these, spin-induced precession and orbital eccentricity play critical roles in shaping the phenomenology of a \ac{gw} signal:
capturing their influence requires sophisticated techniques.

Spin precession occurs when the spin vectors of the binary components are misaligned with the orbital
angular momentum, inducing the precession
of the individual spin vectors $\bm{S}_{1,2}$ and of the orbital angular momentum vector $\bm{L}$ 
around the total angular momentum $\bm{J}$~\cite{Thorne:1984mz, Apostolatos:1994mx,Apostolatos:1996rf}.
Consequently, the orbital plane itself oscillates, introducing characteristic modulations in the \ac{gw} signal. 
The dominant emission occurs in a direction which is approximately that of the Newtonian orbital angular 
momentum~\cite{Schmidt:2010it, Schmidt:2012rh, Boyle:2011gg, OShaughnessy:2011pmr}, which defines a special non-inertial ``co-precessing''
frame where the waveform closely resembles that of an aligned-spin system~\cite{Buonanno:2002fy, Schmidt:2010it, Schmidt:2012rh}. 
State of the art \ac{gw} models track the evolution of this co-precessing frame via three Euler angles, and apply a time-dependent 
rotation (often referred to as the ``twist'') to the co-precessing modes to reconstruct the waveform in a generic inertial reference frame.

Eccentricity (or, more generally, ``non-circularity''), on the other hand, induces oscillations in the frequency and amplitude of the \ac{gw} on the timescale
of the orbit, with each pericenter passage accompanied by  a ``burst'' of \ac{gw} emission.
Additionally, the presence of eccentricity reveals the well-known effect of periastron advance, which
manifests in the waveform as modulations in the amplitude oscillations (see e.g. Fig.~4 of Ref.~\cite{Morras:2025nlp}).
From a modeling perspective, the treatment of non-circular dynamics requires a generalization of the
prescription used to describe \ac{gw} emission and its dissipative effects. Historically, \ac{pn} expressions for the energy and momentum
fluxes have been specialized to \ac{qc} orbits, as eccentricity is efficiently radiated away~\cite{Peters:1963ux}, and thus was expected
to be negligible once the system enters the \ac{lvk} frequency band.
Energy and momentum fluxes for non-circularized binaries have been derived in \ac{pn} theory up to 3 \ac{pn} order (e.g.~\cite{Arun:2007rg, Loutrel:2016cdw, Placidi:2023ofj, Gamboa:2024imd}),
and various examples of their factorizations have been tested in the point-particle limit and against \ac{nr} simulations~\cite{Albanesi:2021rby, Albanesi:2022ywx, Faggioli:2024ugn}.

In this work, we use \TEOBResumS~\cite{Gamba:2024cvy}, a state of the art \ac{eob} model
that combines \ac{pn} and \ac{nr} information to describe coalescing \ac{bbh} systems.
The formalism it is based on, \ac{eob}~\cite{Buonanno:1998gg,Buonanno:2000ef,Damour:2000we,Damour:2001tu,Damour:2008gu,
Damour:2008qf,Barausse:2009xi,Damour:2009sm,Damour:2009wj,Damour:2016gwp,Vines:2017hyw,Damour:2017zjx}, is a resummation of the \ac{pn}
dynamics of the system, including the effects of spins, precession, eccentricity, subdominant modes
and tidal effects.
The conservative Hamiltonian of \TEOBResumS~incorporates point mass information up to 5\ac{pn}
\cite{Bini:2019nra, Nagar:2021xnh}. Spin-orbit contributions are included at 
\ac{nnlo}~\cite{Damour:2014sva,Nagar:2021xnh}, and inverse-Taylor-resummed following the \ac{eob} prescriptions detailed in Ref.~\cite{Nagar:2018zoe}. 
Even-in-spin effects are accounted for up to \ac{nlo}. The radiative sector
contains point-mass and spin terms according to Tab.~I of Ref.~\cite{Nagar:2024oyk}, which also lists their resummation choices.

Within \TEOBResumS, we indirectly define the initial (model) eccentricity $e$ and anomaly $u$ as:
\begin{equation}
  r = \frac{p}{1 + e\cos(u)} \, ,
\end{equation}
that is, the initial separation $r$ is given in terms of the semi-latus rectum $p$ (which can be determined
from the initial average orbital frequency), the eccentricity $e$ and true anomaly $u$, related to the
mean anomaly $\zeta$ via the eccentric anomaly and Kepler's equation~\cite{1973CeMec...7..388B}.
Non-circular effects are then accounted for via the prescription introduced in Ref.~\cite{Chiaramello:2020ehz}: 
the quasi-circular \ac{lo} terms, both 
in the radiation reaction and waveform, are replaced with their exact analytical expressions valid on general orbits.
Note that this simple technique has been validated against a large number of test-mass evolutions, proving to be more reliable with respect
to the direct inclusion of \ac{pn} expanded high order terms.
Additionally, the \textit{radial} radiation reaction force, usually set to zero for quasi-circular binaries, is non-negligible for binaries
coalescing on generic orbits. It is included in the model as a resummation of the 2\ac{pn} expression derived in Ref.~\cite{Bini:2012ji}.
Precession, on the other hand, is modeled via a hybrid \ac{pn}-\ac{eob} scheme. We use the \ac{eob} orbital frequency $\Omega$
to drive the evolution of the orbit-averaged \ac{pn} spin evolution equations of Ref.~\cite{Akcay:2020qrj} up to merger. Beyond this point, identified
as the last peak of the ``pure" orbital \ac{eob} frequency (i.e., the term in $\Omega$'s expression coming from the orbital part of the
\ac{eob} Hamiltonian, excluding the spin-orbit contribution), we implement a \ac{qnm}-inspired prolongation of the Euler angles
that describes the precession of the remnant \ac{bh}'s emission~\cite{Pratten:2020fqn, Gamba:2021ydi}. 
For more details, we refer the reader to Sec.~IID of Ref.~\cite{Gamba:2021ydi}.

\begin{figure}[t]
  \centering
  \includegraphics[width=\textwidth]{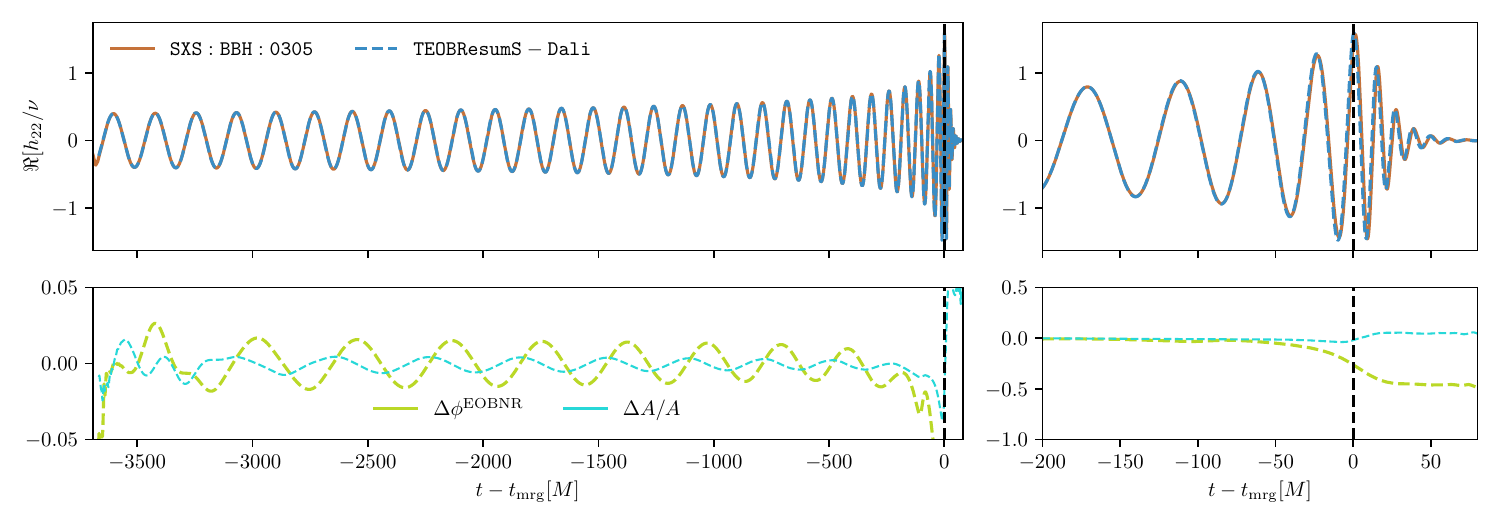}
  \includegraphics[width=\textwidth]{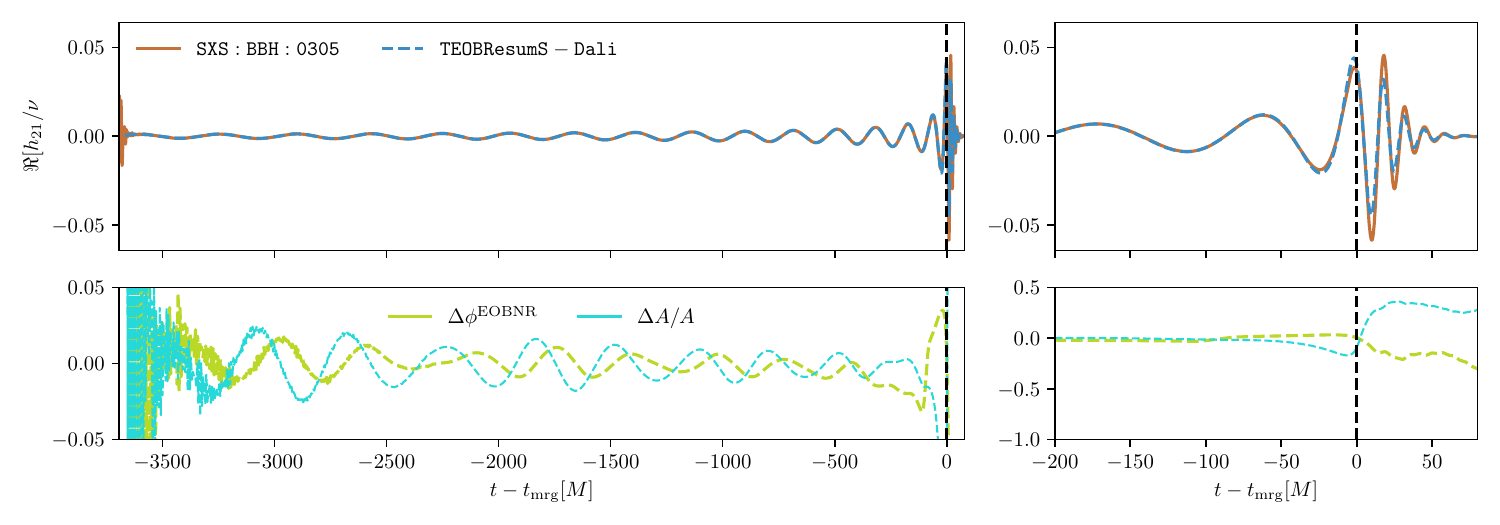}
  \caption{Comparison of the $(2,2)$ and $(2,1)$ modes of the \TEOBResumS~model against the SXS:BBH:0305 simulation,
  which is a \gw-targeted simulation with mass ratio $q=1.22$ and aligned spins $\chi_1=0.33$, $\chi_2=-0.44$.
  The (2,2) mode is aligned in the early inspiral, and the time and phase shifts obtained used also to align the $(2,1)$ mode.
  The small oscillations observed in the EOB/NR phase difference are due to a small residual eccentricity in the \ac{eob} waveform, related to
  the adiabatic initial conditions employed~\cite{Bonino:2022hkj}. The mismatch~\cite{Harry:2017weg} among the two waveforms 
  is below $10^{-3}$ for masses compatible with \gw.
  }
  \label{fig:sxs-305}
\end{figure}

\begin{figure}[t]
\centering
\includegraphics[width=\textwidth]{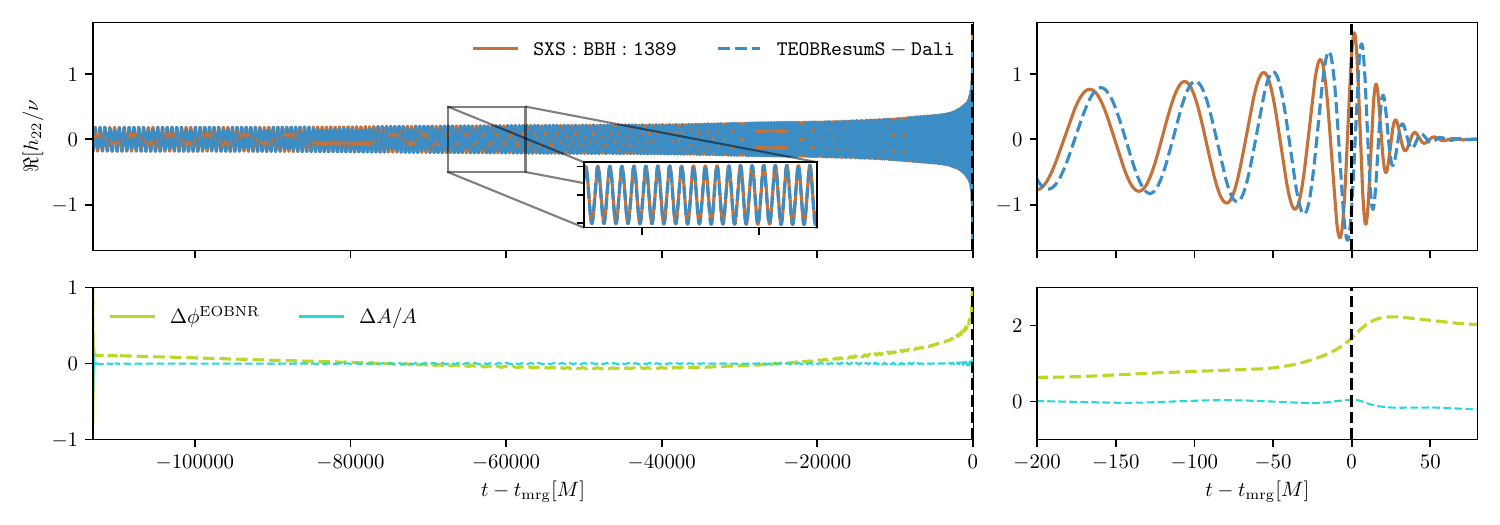}
\includegraphics[width=\textwidth]{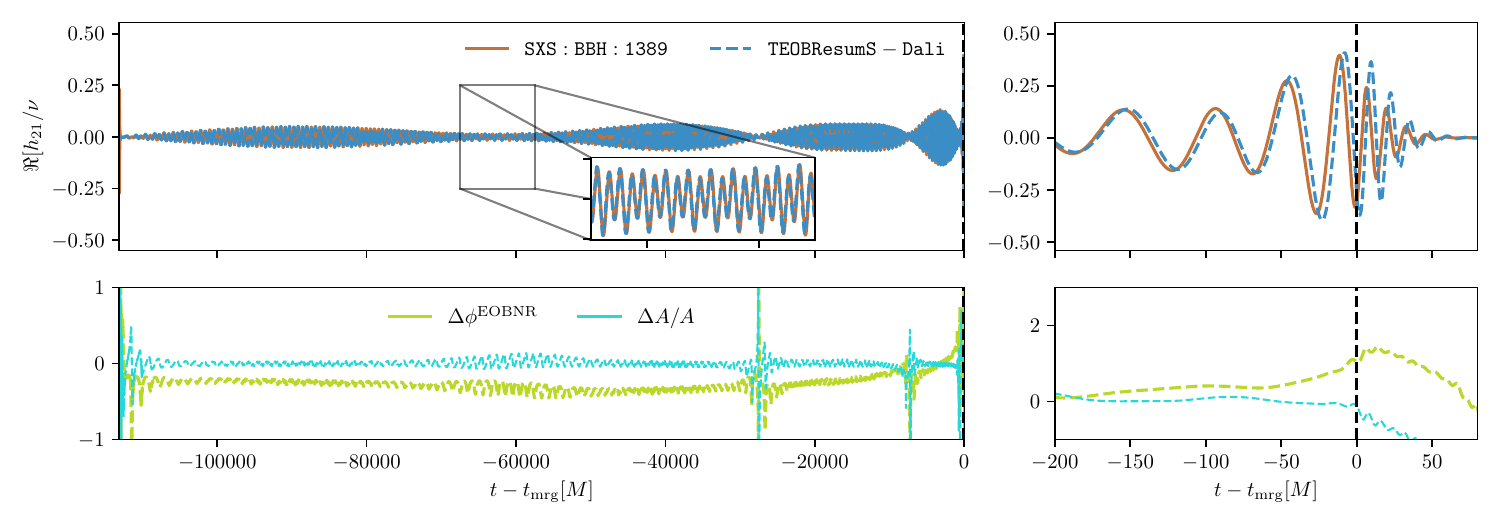}
\caption{Comparison of the $(2,2)$ and $(2,1)$ modes of the \TEOBResumS~model against the SXS:BBH:1389 simulation,
which is a spin-precessing simulation with mass ratio $q = 1.63$, and spins $\bm{\chi}_1 = (-0.29, 0.2,-0.3)$, $ = \bm{\chi}_2 = (-0.1, 0.42, 0.16)$
The $(2,2)$ mode is aligned in the early inspiral, and the time and phase shifts obtained used also to align the $(2,1)$ mode.
Remarkably, the \ac{eob} and \ac{nr} waveforms maintain phase coherence and a good amplitude agreement throughout the entire
inspiral and merger, with the exception of the last few cycles, where the \ac{eob} model exhibits a phase lag of $\sim 1$ rad.
The spin-induced precession of the system is clearly visible in the $(2,1)$ mode: nodes of the amplitude correspond to the
instants at which the azimuthal angle between the inertial frame and the co-precessing frame is equal to $0$.
The mismatch~\cite{Harry:2017weg} among the two waveforms is $\sim 5 \times 10^{-3}$ for masses compatible with \gw.
}
\label{fig:sxs-1389}
\end{figure}

\begin{figure}[t]
 \centering
 \includegraphics[width=\textwidth]{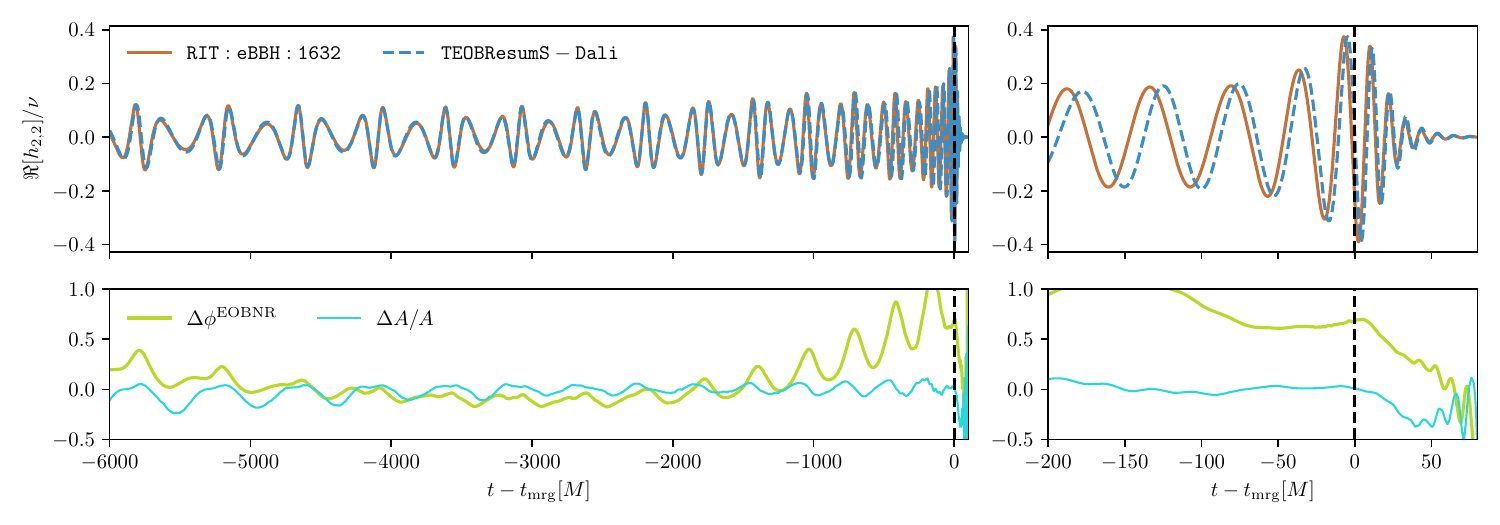}
 \includegraphics[width=\textwidth]{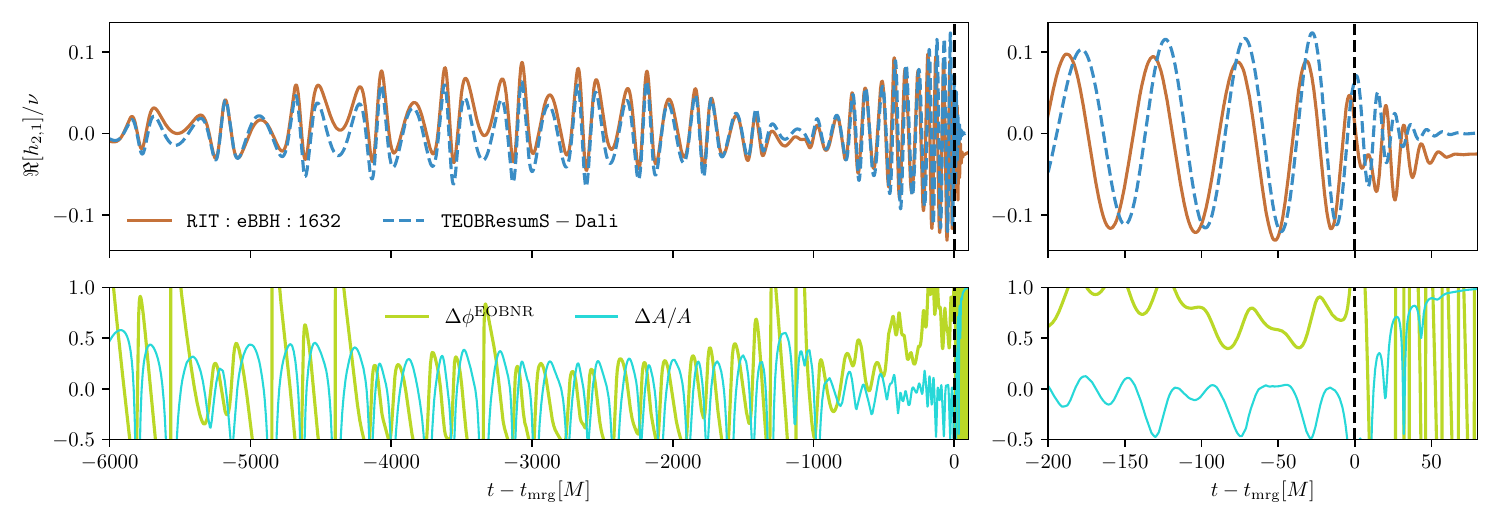}
 \caption{Comparison of the $(2,2)$ and $(2,1)$ modes of the \TEOBResumS~model against the RIT:BBH:1632 simulation,
 which is an eccentric, precessing simulation with mass ratio $q=1$ and aligned spins $\chi_1 = (0.7, 0, 0)$, $\chi_2 = (0.7, 0, 0)$.
 Similarly to the case of SXS:BBH:1389, the $(2,2)$ mode is only weakly affected by the presence of precession. Eccentricity, on the
 other hand, induces very recognizable oscillations in the amplitude and frequency of the waveform associated to the radial motion
 (pericenter passages) of the binary.
 The mismatch~\cite{Harry:2017weg} among the two waveforms is $\mathcal{O}(10{-2})$ for masses compatible with \gw.
 }
\label{fig:rit-1632}
\end{figure}

To both visually illustrate the impact of the various physical effects on the \ac{gw} emission, as well as to display the faithfulness of our model,
we now present three comparisons of \TEOBResumS~with \ac{nr} simulations: 
(i) SXS:BBH:0305~\cite{Lovelace:2016uwp}, which is a \gw-targeted simulation with mass ratio $q=1.22$ and spins $\chi_1=0.33$, $\chi_2=-0.44$ 
aligned with the orbital angular momentum (Fig.~\ref{fig:sxs-305}); 
(ii) SXS:BBH:1389~\cite{Varma:2018mmi}, which is a spin-precessing simulation spanning more than 140 orbits having 
$q=1.63$, $\bm{\chi}_1=(-0.29, 0.2, -0.3)$, $\bm{\chi}_2=(-0.1, 0.42, 0.16)$ (Fig.~\ref{fig:sxs-1389}); 
(iii) RIT:BBH:1632~\cite{Campanelli:2005dd, Nakano:2015pta}, which is an eccentric, precessing \ac{bbh} 
simulation with $q=1$ and $\bm{\chi}_1 = \bm{\chi}_2 = (0.7, 0, 0)$ 
(Fig.~\ref{fig:rit-1632}). For all of the simulations considered, we show the $(2,2)$ and $(2,1)$ modes,
align the waveforms in the early inspiral and compute phase and amplitude differences all through to merger and ringdown.
As expected, the quantitative agreement of our model with \ac{nr} degrades as the complexity of the system increases.
The phase difference at merger $\Delta\phi_{22}^{\rm EOBNR} = \phi_{22}^{\rm EOB} - \phi_{22}^{\rm NR}$, which amounts to $\sim -0.3$ rad for the
quasi-circular, aligned spins system, grows to more than $1$ rad for the spin-precessing and eccentric, precessing cases.
Nonetheless, qualitatively, the main features and oscillations induced by eccentricity and/or precession are fully captured by the model:
from inspecting the $(2,1)$ mode from Fig.~\ref{fig:sxs-1389}, we immediately observe that the four precession cycles that the system undergoes
(as inferred from the number of minima of the amplitude of this mode) are reproduced by the simple prescription employed.
Similarly, periastron passages (corresponding to the amplitude peaks of Fig.~\ref{fig:rit-1632}) are correctly captured as well.

A quantitative and comprehensive validation of our model against \ac{nr} simulations has been performed
in Ref.~\cite{Albanesi:2025txj}. There, \TEOBResumS~was compared with $1395$ \ac{nr} simulations from the
SXS~\cite{Boyle:2019kee}, RIT~\cite{Healy:2020vre,Healy:2022wdn}, CoRe~\cite{Dietrich:2018phi,Gonzalez:2022mgo} and 
ICC catalogs~\cite{Andrade:2023trh}, 
covering a wide range of masses, spins and orbital configurations.
Anticipating the results of Sec.~\ref{sec:pe}, we find that in the region of interest for \gw~(that is, equal-mass, slowly spinning \acp{bbh})
our model is typically more than $99.9\%$ faithful to \ac{nr}.
The high eccentricity regime has been further tested in Ref.~\cite{Bhaumik:2024cec}, where the model was shown to be
in agreement with \ac{nr} simulations with initial eccentricities as high as $e \sim 0.7$.

\section{GW parameter estimation}
\label{sec:pe}

Given some data $d$ collected by the detectors, the goal of \ac{gw} \ac{pe} is to estimate the
probability distribution of the parameters $\bm{\theta}$ which, under some hypothesis $H$, describe a binary system given the data, $p(\bm{\theta}|d, H)$.
This problem is typically tackled with the use of Bayesian inference
techniques~\cite{Veitch:2014wba, Ashton:2018jfp, Breschi:2021wzr}, according to which $p(\bm{\theta}|d, H)$ can be estimated via:
\begin{equation}
  p(\bm{\theta}|d, H) = \frac{p(d|\bm{\theta}, H)p(\bm{\theta}| H)}{p(d | H)}.
\end{equation}
In the expression above, $p(d|\bm{\theta}, H)$ is the likelihood function, which
contains the information about the \ac{gw} signal and the noise
of the detectors;
$p(\bm{\theta} | H)$ is the prior distribution, which encodes
our assumptions about the distribution of the parameters before the data is observed, and is usually chosen to be
uniform in the parameters of interest (``agnostic'' prior), or motivated by physical models;
$p(d|H) = \int d\bm{\theta} p(d|\bm{\theta}, H) p(\bm{\theta}| H)$ is the evidence (or ``marginal likelihood''), a constant that ensures the posterior is properly normalized. 
In the context of model selection, $p(d | H)$ is used to compute the \ac{bf} between two models, 
which quantifies the relative probability of the data given the two models and is used to determine 
whether one is favored by the data:
\begin{equation}
  \log\mathcal{B}^{1}_2 = \log p(d|H_1) - \log p(d|H_2) = \log \frac{p(d|H_1)}{p(d|H_2)}.
\end{equation}

We apply this framework to \gw. We describe the system via a set of 17 parameters: 
the masses of the binary components $m_1$ and $m_2$, the
(dimensionless) spin vectors of the components $\bm{\chi_1}$ and $\bm{\chi_2}$, the initial eccentricity
of the orbit $e$ and the mean anomaly of the system $\zeta$, the luminosity distance of
the source to the detectors $D_L$, the inclination of the orbit with respect to the line of sight
$\iota$, the polarization angle $\psi$, the time of coalescence $t_c$ at which the signal
reaches its peak amplitude\footnote{More precisely, we use the \textit{last} peak of the waveform
invariant amplitude to define merger, see e.g. App.~C of Ref.~\cite{Albanesi:2024xus}}, the reference orbital phase $\phi_{\rm ref}$, and the
sky position of the source $(\alpha, \delta)$.
We sample in chirp mass $\mathcal{M} \in [20, 40] M_\odot$ and (inverse) mass ratio $1/q \in [0.05, 1]$, enforcing uniform priors on the component masses $m_1, m_2$,
constrained to be $\in [3, 100] M_\odot$;
over the whole sky area for the angles $(\alpha, \delta)$, impose
uniform priors with periodic boundaries on $\psi \in [0, 2\pi]$ and $\phi_{\rm ref} \in [0, 2\pi]$, and employ
a power-law prior for the luminosity distance $D_L\in [10, 3000]$ Mpc. 
We use a flat prior for the time of coalescence $t_c \in [t_0-1.6\mathrm{s}, t_0+1.6\mathrm{s}]$, where the reference 
GPS time is $t_0=1126259462.891$, but instead of sampling it we analytically marginalize over it when computing the likelihood.
Rather than directly sampling the spin components, we decompose the spin vectors in terms of their magnitudes,
their tilt angles with respect to the orbital angular momentum $\bm{L}$,
the azimuthal angle separating the spin vectors $\phi_{12}$ and the azimuthal angle angle $\phi_{\rm JL}$.
Further, rather than directly sampling in $\iota$, we sample the cosine of the angle $\theta_{\rm JN}$
between the line of sight and the total angular momentum $\bm{J}$ of the system.
Given our aim of contrasting four different scenarios, each corresponding to a different orbital
configuration, priors on $e, \zeta, \bm{\chi_1}$ and $\bm{\chi_2}$ depend on the chosen underlying hypothesis:
\begin{itemize}
  \item \textit{quasi-circular, aligned-spin case (``QC-AS'')}: we fix $e=0, \zeta=0$ and sample the spins using an ``aligned spin'' prior, considering
  magnitudes $|\bm{\chi}_1| \in [0, 0.99]$,  $|\bm{\chi}_2| \in [0, 0.99]$. The full model is thus $10$-dimensional.
  \item \textit{Quasi-circular, precessing-spin case (``QC-PS'')}: we fix $e=0, \zeta=0$ and use an isotropic spin prior. The full model is thus $14$-dimensional.
  \item \textit{Non-circular, aligned-spin case (``NC-AS'')}: we use uniform priors on $e \in [0, 0.4]$\footnote{These prior bounds are inspired by previous analyses of
  the event, which found it to be consistent with a low eccentricity merger~\cite{Bonino:2022hkj, Ramos-Buades:2023yhy, Gamboa:2024hli}} and $\zeta \in [0, 2\pi]$, and
  sample over the spin parameters employing the same prior as the QC-AS case. The full model is thus $12$-dimensional.
  \item \textit{Non-circular, precessing-spin case (``NC-PS'')}: we use uniform priors on $e \in [0, 0.4]$ and $\zeta \in [0, 2\pi]$, and
  an isotropic spin prior. The full model is thus $16$-dimensional.
\end{itemize}

We evaluate the posterior distribution $p(\bm{\theta} | d)$ using the 
\texttt{bilby} package~\cite{Ashton:2018jfp}, coupled with the nested sampling algorithm as
implemented in \texttt{dynesty}~\cite{Speagle:2020}. We use $2048$ livepoints and $60$ as number of accepted steps 
(\texttt{naccept}), running the analysis on $48$ cores.
We analyze $8$ seconds of data around the GPS time of the event, focusing on the frequencies 
between $20$ and $896$ Hz. We download the data from the \ac{gwosc} and
use the publicly available \ac{psd} and calibration envelopes as computed by the 
\ac{lvk} collaboration~\cite{LIGOScientific:2018mvr}.
We generate waveforms with \TEOBResumS~using a sampling rate of $4096$ Hz, and
consider the $(\ell, |m|) = (2,1), (2,2), (3,3)$ and $(4,4)$ co-precessing modes for the waveform generation.
We use an initial orbit-averaged frequency for waveform generation of $15$ Hz, lower than the minimum one in the analysis. 
This choice is motivated by the fact that the instantaneous frequency of an eccentric waveform varies throughout each orbit. 
Therefore, if we were to generate waveforms starting at $20$Hz we would risk miss some signal power.
As our model natively outputs waveforms in the time domain, each likelihood evaluation requires us to perform a \ac{fft}.
With these settings, the analyses take between $9$ and $17$ days to complete, depending on the model
complexity.

\begin{table}[h!]
  \centering
  \begin{tabular}{|c|c|c|c|c|c|}
          \hline
          Parameter & GWTC & QC AS & QC PS & NC AS & NC PS \\
          \hline
          $\mathcal{M} [M_\odot]$& $30.4^{+1.6}_{-1.7}$& $30.9^{+1.4}_{-1.4}$& $31.2^{+1.6}_{-1.6}$& $30.6^{+1.5}_{-1.5}$& $30.9^{+1.6}_{-1.6}$ \\
          $1/q$& $0.85^{+0.14}_{-0.22}$& $0.89^{+0.10}_{-0.17}$& $0.87^{+0.12}_{-0.22}$& $0.89^{+0.10}_{-0.17}$& $0.86^{+0.13}_{-0.21}$ \\
          $\chi_{\rm eff}$& $-0.05^{+0.11}_{-0.15}$& $-0.03^{+0.10}_{-0.12}$& $-0.01^{+0.12}_{-0.13}$& $-0.06^{+0.11}_{-0.12}$& $-0.03^{+0.12}_{-0.13}$ \\
          $d_L$ [Mpc] & $460^{+130}_{-140}$& $440^{+160}_{-160}$& $480^{+140}_{-150}$& $430^{+160}_{-160}$& $470^{+130}_{-150}$ \\
          $\iota$ [rad]& $2.63^{+0.36}_{-0.43}$& $2.61^{+0.38}_{-0.56}$& $2.79^{+0.26}_{-0.54}$& $2.61^{+0.39}_{-0.57}$& $2.80^{+0.26}_{-0.56}$ \\
          $e$& -- &  -- & -- & $0.04^{+0.06}_{-0.04}$& $0.04^{+0.06}_{-0.04}$ \\
          $\zeta$ [rad] & -- & -- & -- & $3.3^{+2.7}_{-2.9}$& $3.3^{+2.7}_{-3.0}$ \\
          $\chi_{\rm p}$& $0.51^{+0.38}_{-0.40}$& -- & $0.41^{+0.41}_{-0.32}$& -- & $0.40^{+0.42}_{-0.31}$ \\
          $\log{\mathcal{B}}^{S}_{N}$ & -- & $280.55^{+0.11}_{-0.11}$ & $280.44^{+0.11}_{-0.11}$ & $279.41^{+ 0.12}_{- 0.12}$ & $279.29^{+0.12}_{-0.12}$ \\
          \hline
  \end{tabular}
  \caption{Summary of the parameters of \gw~from GWTC-2.1 (first column) and under the four hypotheses considered in this work.
  The (natural) logarithm of the \ac{bf} 
    comparing the signal to the noise hypothesis $\log \mathcal{B}^S_N$
    is reported in the last row, confirming the expectation that this event is most consistent with a
  quasi-circular inspiral of two \acp{bh}, and that non-circularity is disfavoured.}
  \label{tab:summary}
\end{table}

\begin{figure}[t]
  \centering
  \includegraphics[width=\textwidth]{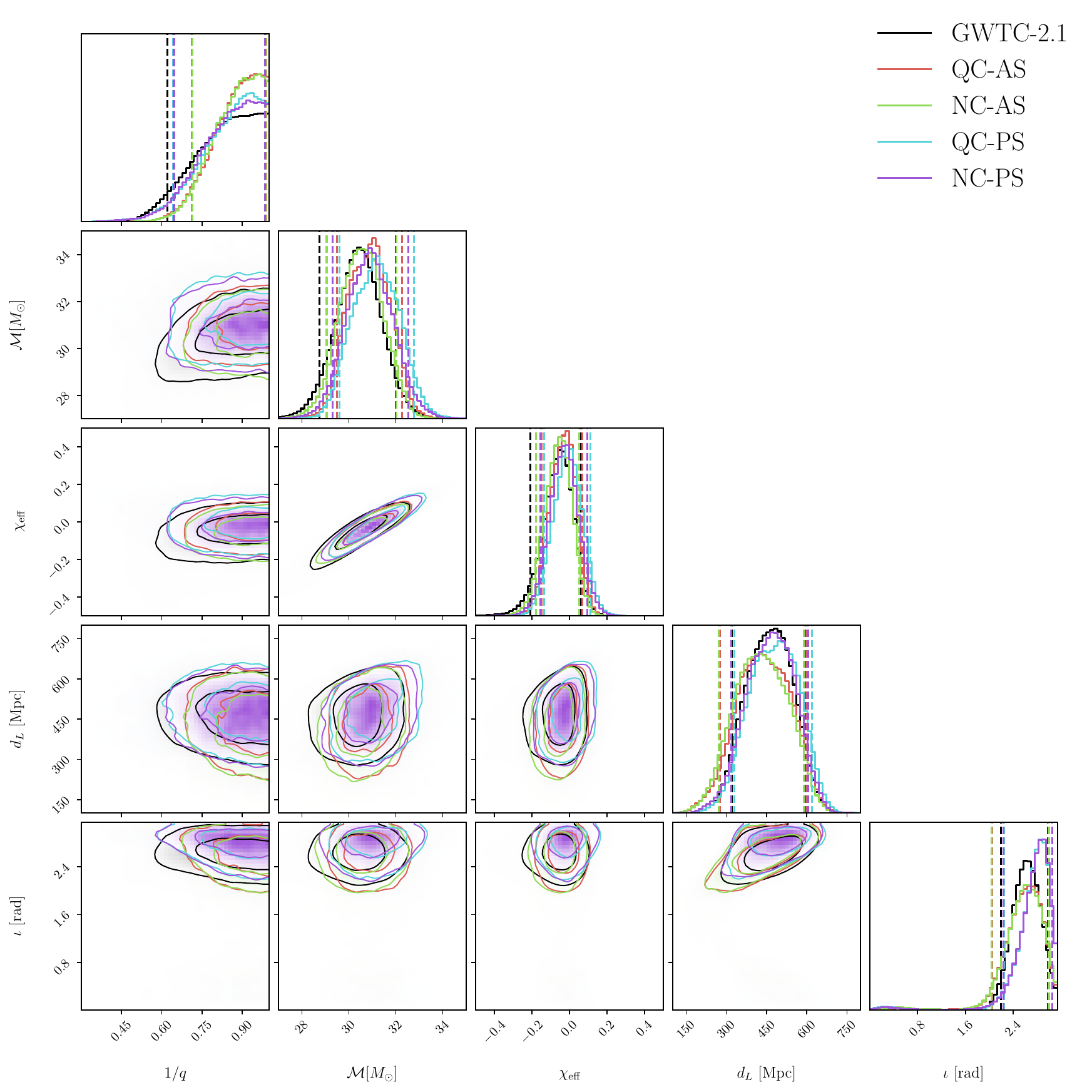}
  \caption{Marginalized two- and one- dimensional posterior distributions for the common parameters of \gw~in the four scenarios considered.
    We additionally compare our results to those from the GWTC catalog paper~\cite{LIGOScientific:2021usb},
    here shown as the black contours. We find broad agreement between all analyses, with GWTC-2.1 predicting slightly lower
    chirp mass and inverse mass ratio}
  \label{fig:post_common}
\end{figure}

\section{Results}
\label{sec:results}

\begin{figure}[t]
  \centering
  \includegraphics[width=0.45\textwidth]{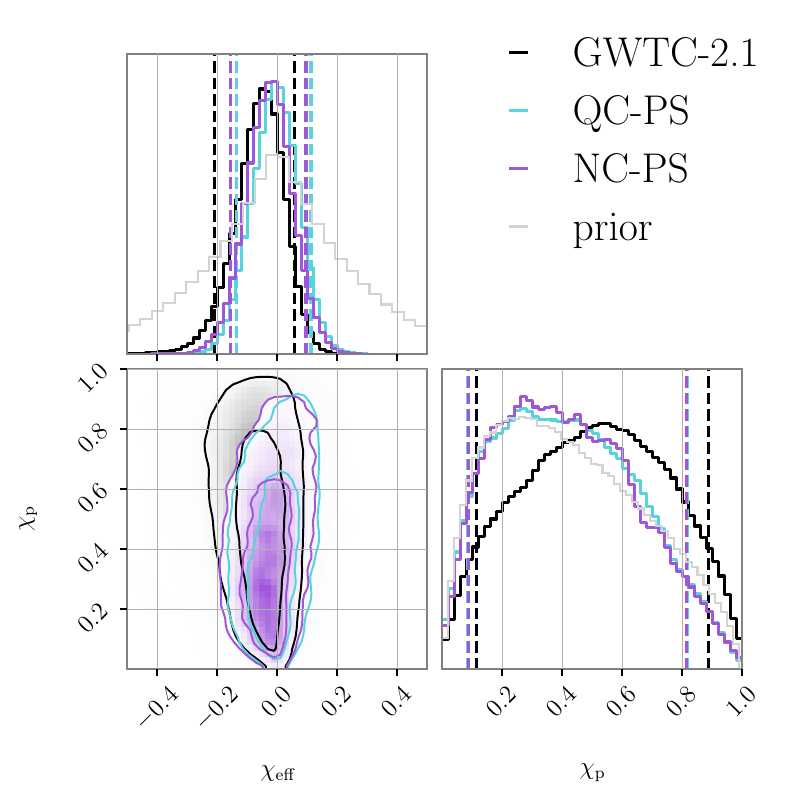}
  \includegraphics[width=0.45\textwidth]{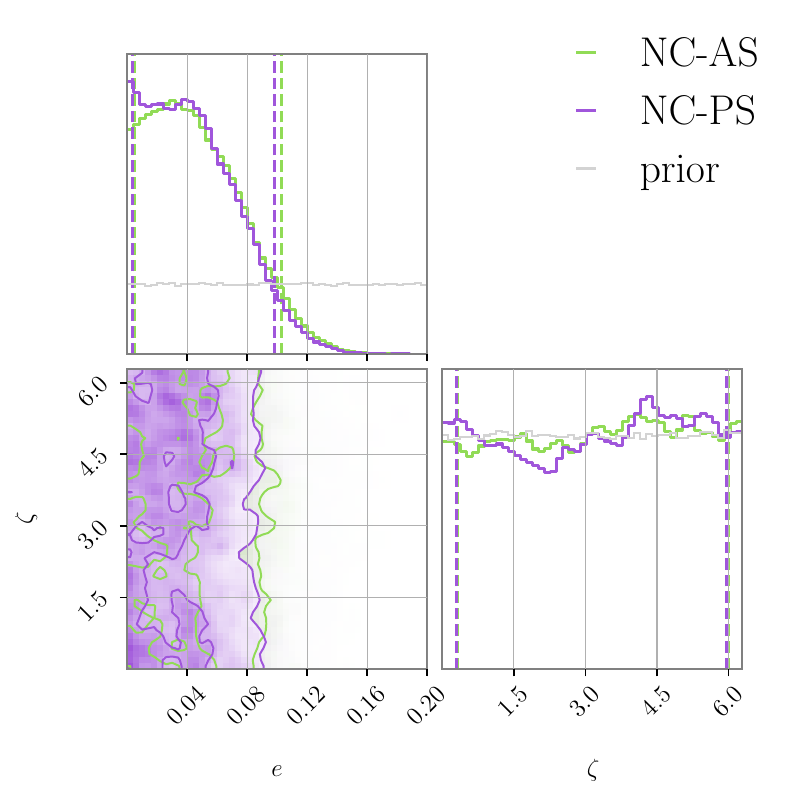}
  \caption{Marginalized two- and one- dimensional posterior distributions for the spin parameters $\chi_{\rm eff}$ and $\chi_p$ (left) and
    the eccentricity $e$ and mean anomaly $\zeta$ (right). While $\chi_{\rm eff}$ is well constrained and in all case consistent with zero, 
    $\chi_p$ spans the entire prior. The eccentricity at $15$ Hz is well constrained, with $e < 0.08$ at $90\%$ credibility. Finally, the mean anomaly
    is follows the uniform prior employed. This is consistent with the expectation of a quasi-circular binary, where no value of $\zeta$ is preferred.
    }
  \label{fig:spins-ecc}
\end{figure}

Fig.~\ref{fig:post_common} shows the posterior distributions for the common intrinsic parameters of
\gw~in the four scenarios considered. Most notably, the masses and effective spin are broadly
consistent among all analyses, with $\mathcal{M} = 30.90^{+1.63}_{-1.59}$,
$1/q = 0.86^{+0.13}_{-0.21}$ and $\chi_{\rm eff} = -0.03^{+0.12}_{-0.13}$ for the non-circular, precessing-spin model.
The values recovered are also in agreement with those reported by past studies performed with 
different models~\cite{Pratten:2020ceb, Estelles:2020twz, Ramos-Buades:2023ehm, Gamboa:2024hli},
as well as from GWTC-2.1~\cite{LIGOScientific:2021usb}.
The nominal model eccentricity (left panel of Fig.~\ref{fig:spins-ecc}) is constrained for both the planar and precessing spins
scenarios to $e = 0.04^{+0.06}_{-0.04}$ at $90\%$ credibility at $15$ Hz,
showing considerable support towards the left boundary $e=0$. The anomaly parameter, instead, cannot be
measured and closely follows the flat prior employed, confirming the expectation of a quasi-circular binary.
Finally, in-plane spin components (right panel of Fig.~\ref{fig:spins-ecc}) are also not well constrained,
with $\chi_{\rm p} = 0.40^{+0.42}_{-0.31}$ spanning the width of the entire prior distribution.
With respect to the distrubution reported in GWTC-2.1, we find a slight shift towards lower values of $\chi_{\rm p}$,
which is however consistent with, e.g., the results of Ref.~\cite{Ramos-Buades:2023ehm} and points
towards small differences due to the modeling of spin precession.

\begin{figure}[t]
  \centering
  \includegraphics[width=\textwidth]{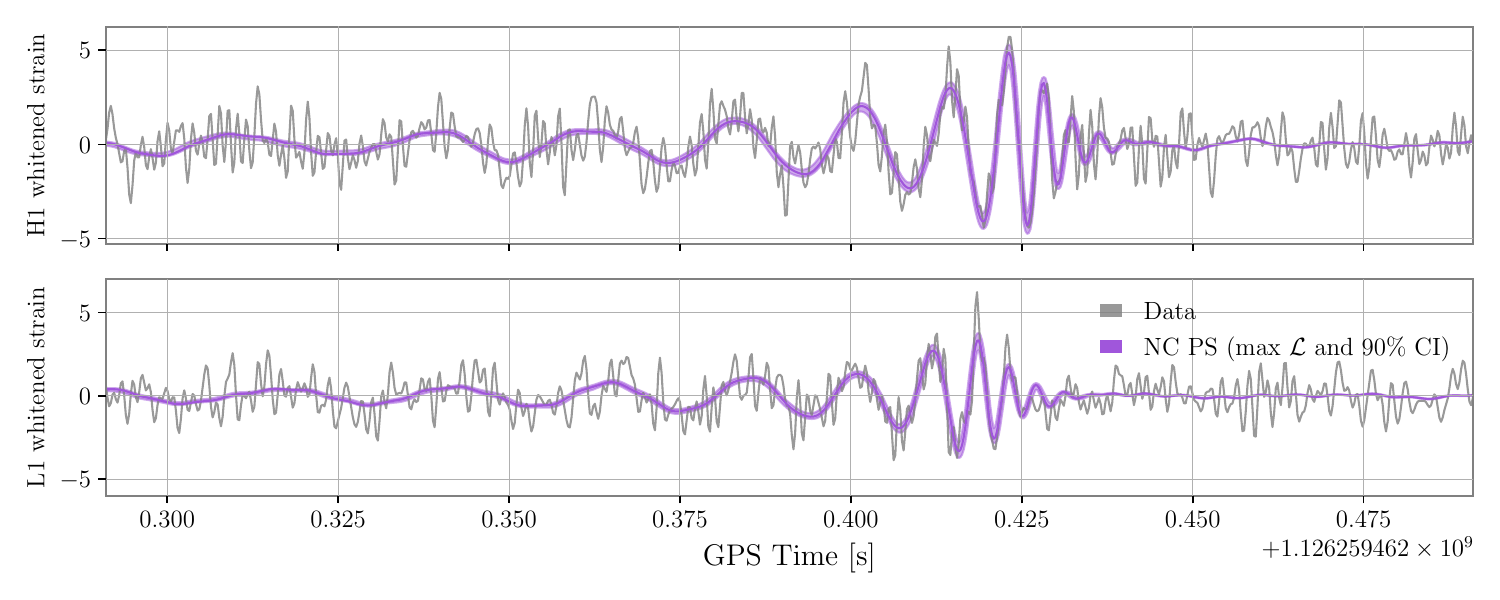}
  \caption{Whitened strain data from the Hanford (top) and Livingston (bottom) detectors,
    with the prediction from the non-circular, precessing-spin model overlaid; the thick line corresponds to the maximum likelihood waveform, the colored region is the 90\% equal-tailed credible interval.
    As can be seen, the signal is visually consistent with a \ac{qc} \ac{bbh} merger, with no
    clear evidence of eccentricity or precession. 
      This is reflected in the fact that the bulk of the posterior distribution for this model has 
      low effective and precessing spin parameters, and eccentricity consistent with zero.
    }
  \label{fig:reconstructed}
\end{figure}

When comparing the different hypotheses in terms of log Bayes factors,
we do not find any scenario to be strongly favored over the others (see Tab.~\ref{tab:summary}).
However, we do find that adding non-circularity to the model tends to decrease its Bayesian evidence,
with the QC-AS and QC-PS models being favored over their NC counterparts by $\log\mathcal{B}^{\rm QC}_{\rm NC} \sim 1$.
As such, we conclude that the data is consistent with a quasi-circular \ac{bbh} merger, with
small effective spin and no strong evidence in favor of precession. To visually confirm our result, 
we show the time-domain reconstruction from the
non-circular, precessing-spin model in Fig.~\ref{fig:reconstructed}, overlaid on the whitened strain data
from the Hanford and Livingston detectors. Compared to the simulations considered in the previous section, the
(whitened) reconstructed waveforms show no clear signatures of eccentricity or precession.

\section{Conclusions}
\label{sec:conclusions}
Despite the fact that \gw~has been the subject of extensive research since its discovery,
its usefulness as a testbed for new techniques and models has not diminished throughout the years.
Its re-analysis as presented in this work provides a significant step forward in the field of \ac{gw} astrophysics.
First, the findings presented a-posteriori validate a considerable body of past research, performed assuming that \gw~is a quasi-circular \ac{bbh} merger.
Second, they represent the first instance of a full \ac{gw} \ac{pe} performed with an inspiral-to-ringdown non-planar, non-circular model. 
This achievement underscores the growing sophistication of \ac{gw} modeling and \ac{pe} techniques, 
highlighting the importance of incorporating more realistic, complex waveforms when interpreting observational data.

Nonetheless, more work is needed to further improve \ac{gw} models in view of
next generation space- and ground- based detectors. Beyond improvements
in their accuracy, additional physical effects
should be included in the waveform generation. These include higher order harmonics beyond those 
considered here, including $m=0$ modes which encode \ac{gw}
memory~\cite{Blanchet:1992br, 1991PhRvL..67.1486C, Favata:2010zu, Rossello-Sastre:2024zlr, Albanesi:2024fts, Rossello-Sastre:2025gtq, Rossello-Sastre:2025dep},
improved descriptions of tidal heating and torquing~\cite{Mukherjee:2022wws, Munna:2023vds, Chiaramello:2024unv}, as well as phenomenological
descriptions of environmental effects~\cite{Barausse:2014pra, Barausse:2014tra, Kavanagh:2020cfn}. The latter, while not expected to be relevant
for \gw~\cite{CanevaSantoro:2023aol}, may be important for the analysis of future events, especially those
originating from dense stellar environments~\cite{Toubiana:2020drf, Speri:2022upm, Roy:2024rhe, Garg:2024qxq, Romero-Shaw:2024klf, DeLuca:2025bph}

The increased computational cost of \TEOBResumS~due to the inclusion of both eccentricity and precession
is a clear limitation for the analysis of large catalogs of \ac{gw} events. Various avenues
can be explored to address this issue.
On the waveform generation side, the use of surrogate models~\cite{Field:2013cfa, Blackman:2015pia, Varma:2018mmi, Varma:2019csw, 
Schmidt:2020yuu, Tissino:2022thn, Islam:2025llx}
can significantly speed up the generation of the waveforms. Preliminary results in this
direction have been presented, for nonspinning~\cite{Barta:2018,Islam:2021} and aligned-spin~\cite{Yun:2021,Shi:2024} eccentric binaries. Recently, a surrogate model for aligned-spin 
\TEOBResumS~was constructed~\cite{Islam:2025llx}, employing a further decomposition of each harmonic mode of emission into modes with monotonically increasing frequency.
Alternatively, reduced order modeling techniques~\cite{Field:2013cfa, Smith:2016qas, Qi:2020lfr, Gadre:2022sed, Tissino:2022thn} 
or analytical approximations~\cite{Damour:2004bz, Cho:2021oai, Klein:2018ybm} can also be employed to speed up the generation of the waveforms.
On the \ac{pe} side, the use of more efficient sampling algorithms~\cite{Williams:2021qyt,Karamanis:2022}, 
deep-learning techniques~\cite{Green:2020dnx, Dax:2021tsq} used as high-quality posterior proposals~\cite{Dax:2022pxd} or approximate or marginalized
likelihoods~\cite{Pankow:2015cra, Lange:2018pyp, Roulet:2022kot} can significantly reduce the computational cost of the analysis; for a review, see Ref.~\cite{Roulet:2024}.

The ability to perform \ac{pe} on systems where orbital eccentricity and spin precession
play significant roles opens the door to more detailed and comprehensive studies of the \ac{bbh} population,
as the two effects are closely linked to the formation channels of \acp{bh} in dynamical environments.
For instance, \acp{bbh} formed in galactic fields are generally expected to exhibit spins aligned with the
orbital angular momentum and to inspiral along quasi-circular
orbits~\cite{Kalogera:1999tq, Belczynski:2017gds, Stevenson:2017tfq, Zaldarriaga:2017qkw, Gerosa:2018wbw}.
In contrast, \acp{bbh} originating from dense stellar environments are more likely to display
significant spin misalignments and residual eccentricities due to frequent dynamical
interactions~\cite{PortegiesZwart:2002iks, Antonini:2016gqe, Rodriguez:2019huv, Gerosa:2021mno}. 
As gravitational-wave detectors continue to increase in sensitivity, future analyses utilizing waveform
models such as \TEOBResumS~will provide deeper insights into the properties of \ac{bbh} systems,
shedding light on their formation pathways and evolutionary histories.

\section*{Acknowledgments}
The authors would like to thank S.~Bernuzzi and A.~Nagar for 
useful discussions. RG is grateful to K.~Chandra and I.~Gupta for constant
support and encouragement throughout the development of this work, and
to M.~Poulsen, J.~Larsen and K.B.~Larsen for inspiration.
The version of \TEOBResumS~employed in this work is available at 
\url{https://bitbucket.org/teobresums/teobresums/src/2505.21612/}.
RG acknowledges support from NSF Grant PHY-2020275
(Network for Neutrinos, Nuclear Astrophysics, and Symmetries (N3AS)).
JL and DC acknowledge support from the Italian Ministry of University and Research (MUR) via the PRIN 2022ZHYFA2, 
{\it GRavitational wavEform models for coalescing compAct binaries with eccenTricity} (GREAT).
ST acknowledges support from NSF grant NSF PHY-2207780.
This material is based upon work supported by NSF's LIGO Laboratory which is a major facility
fully funded by the National Science Foundation.
The authors are grateful for computational resources provided by LIGO Laboratory and supported by
National Science Foundation Grants PHY-0757058 and PHY-0823459.
\section*{References}
\providecommand{\newblock}{}

\end{document}